\documentclass[epjCONF]{svjour}
\usepackage{graphics}
\usepackage[varg]{txfonts} 
\usepackage[latin1]{inputenc}

\newcommand{\be}{\begin{equation}}
\newcommand{\ee}{\end{equation}}
\newcommand{\bea}{\begin{eqnarray}}
\newcommand{\eea}{\end{eqnarray}}
\newcommand{\bean}{\begin{eqnarray*}}
\newcommand{\eean}{\end{eqnarray*}}
\newcommand{\vecp}{{\mathbf p}}
\newcommand{\md}{\mathrm{d}}
\newcommand{\vecx}{{\mathbf x}}
\newcommand{\vecnull}{{\mathbf 0}}

\newcommand{\Dp}{D^{+}}
\newcommand{\Dm}{D^{-}}

\session-title{Hadronic resonance production in heavy ion and elementary collisions}
\begin{document}
\title{In-medium hadron properties from lattice QCD}
\author{Heng-Tong Ding\inst{1}\fnmsep\thanks{\email{htding@quark.phy.bnl.gov}}}
\institute{Physics Department, Brookhaven National Laboratory, Upton, NY 11973, USA }
\abstract{
I review recent results from lattice QCD calculations on the in-medium hadron properties. I discuss the
thermal dilepton rates, heavy quarkonium properties as well as the chiral and $U(1)_A$ symmetries at finite temperature. 
 } 
\maketitle
%

\section{Introduction}
\label{intro}

Understanding the phases and properties of  strongly interacting matter at high temperature and
density  has been one of the central goals of research in nuclear science. The hadrons inside the medium 
would feel the modifications when the medium is heated or compressed~\cite{Rapp:2011zz}. The vector meson $\rho$ and the axial vector
meson $a_1$ are degenerate when the chiral symmetry is restored from the spontaneously broken phase at a certain temperature.  The particles
$\eta^{\prime}$ and $\eta$ are degenerate when the $U(1)_A$ symmetry is restored. The heavy quarkonia, e.g. $J/\psi$ and $\Upsilon$,
due to their small size and large mass, may survive up to certain critical temperatures and could serve as medium thermometers~\cite{Matsui:1986dk}.
The thermal modifications of hadrons are encoded in the hadron spectral functions, e.g. the presence (absence) of 
a resonance peak in the spectral function indicates the existence (dissociation) of a bound state. The hadron spectral function
is also related to thermal dilepton and photons rates as well as transport properties of the medium, e.g. the electrical conductivity and heavy quark
diffusion coefficient. The related hadron correlation functions and their susceptibilities are useful observables that allow to study the chiral phase transition 
and the $U(1)_A$ symmetry restoration.

In the following I will give brief introductions to meson correlation and spectral functions in section~\ref{sec:basic_spf} , and to the chiral 
and $U(1)_A$ symmetries in section~\ref{sec:basic_symmetry}. In section~\ref{sec:results} I will show the recent lattice QCD results on in-medium hadron properties. 
For a review on earlier results see recent reviews, e.g. Ref.~\cite{DeTar:2009ef}. Finally I will give a
summary in section~\ref{sec:sum}.

\section{Meson correlators and spectral functions}
\label{sec:basic_spf}

In-medium meson properties as well as some transport coefficients are all encoded in thermal meson spectral functions. 
The meson spectral function $\rho_H(\omega,\vecp)$  for a given 
quantum number channel $H$ can be defined through the retarded correlator $D_{H}^{R}(\omega,\vecp)$
\bea
\rho_H(\omega,\vecp) = \Dp_H(\omega,\vecp) - \Dm_H(\omega,\vecp) = 2\,{\rm Im}D^R_H(\omega,\vecp),
                                         \label{spf_definition_p}
\eea
where $D^{+(-)}_H(\omega,\vecp)= \int \md^4x\,e^{i\omega t -i\vecp\vecx} \,D_{H}^{+(-)}(t,\vecx)$, $i\Dp_H(t,\vecx) = \left\langle J_H(t,\vecx) J_H(0,\vecnull) \right\rangle$ 
and $i \Dm_H(t,\vecx) = \left\langle J_H(0,\vecnull) J_H(t,\vecx) \right\rangle$.
The local meson operator $ J_{H}(t,\vecx)$ has  the form of
$ J_{H}(t,\vecx)=\bar{\psi}(t,\vecx)~\Gamma_{H} ~\psi(t,\vecx)$ ,
  with $\Gamma_{H}=1,\gamma_5,\gamma_{\mu},\gamma_5\gamma_{\mu}$, for scalar ($SC$), pseudo scalar ($PS$), vector ($VC$) and axial vector ($AV$) channels, respectively. The relation of these quantum numbers to different meson states  is summarized in Table \ref{table:meson_states}.

\begin{table}[htb]
\centering
\scalebox{0.9}{
\begin{tabular}{|c|c| c|c|c|c|c|c|}
\hline
${\rm Channel}$ &$\Gamma_{H}$ & $^{2S+1}L_{J}$  & $J^{PC}$  & $u\bar{u}$ states &$c\bar{c}$ states &$b\bar{b}$ states\\
\hline 
${\rm PS}$         & $\gamma_{5}$      &     $^{1}S_{0}$   &  $0^{-+}$ &  $\pi$ &$\eta_{c}$&  $\eta_{b}$\\
${\rm VC}$          & $\gamma_{\mu} $   &     $^{3}S_{1}$   &  $1^{--}$   & $\rho$     &  $J/\psi$   &$\Upsilon$  \\
${\rm SC}$          & 1                 &     $^{3}P_{0}$   &  $0^{++}$   & $a_0$ &$\chi_{c0}$              &  $\chi_{b0}$     \\
${\rm AV}$         & $\gamma_{5}\gamma_{\mu}$     &     $^{3}P_{1}$   &  $1^{++}$ &  $a_1$ &$\chi_{c1}$ &$\chi_{b1}$\\ 
\hline
\end{tabular}
}
\caption{Meson states in different channels for light, charm and bottom quarks.}
\label{table:meson_states}
\end{table}

 The spectral function in the vector channel $\rho_V(\omega,\vecp,T)$ is related to the experimentally accessible differential cross section for thermal dilepton production,
\be
\frac{\md W}{\md\omega\,\md^3\vecp} = \frac{5\alpha_{em}^2}{54\pi^3}\frac{1}{\omega^2(e^{\omega/T} -1)}\,\rho_V(\omega,\vecp,T),
\ee
where $\alpha_{em}$ is the electromagnetic fine structure constant. 
The low frequency limit of the vector spectral function is related to the transport properties of the medium. In the light meson case,
the spatial components of the vector spectral function yield the electrical conductivity

\be
\frac{\sigma}{T} = \frac{C_{em}}{6} \lim_{\omega \rightarrow 0} \lim_{\vecp \rightarrow 0} 
\frac{\rho_{ii}(\omega,\vecp, T)}{\omega T} \; ,
\label{conduct}
\ee
where $C_{em}$ is the sum of the square of the elementary charges of the quark flavor $f$, $C_{em}=\sum_f Q_f^2$.
The electrical conductivity quantifies the response of the QGP to electromagnetic fields and is also related to the 
emission rate of soft photons, 
\be
\lim_{\omega \rightarrow 0} \omega \frac{{\rm d} R_\gamma}{{\rm d}^3p} =
\frac{3}{2\pi^2}\, \sigma(T) \,T \,\alpha_{em} \ .
\label{softphoton}
\ee
Furthermore, the photon emission rate of a thermal medium  $\omega \frac{{\rm d} R_\gamma}{{\rm d}^3p}$ is also related to the light vector spectral function at light-like 4-momentum 
 \be
\omega \frac{{\rm d} R_\gamma}{{\rm d}^3p} =C_{em}\, \frac{\alpha_{em}}{4\pi^2} \,
\frac{\rho_{V}(\omega =|\vec{p}|, T)}{{\rm e}^{\omega/T} -1} \ .
\label{photon}
\ee
The spatial components of  the heavy quark vector spectral function are related to the heavy quark diffusion coefficient $D$
\be
D = \frac{1}{6\chi^{00}}\lim_{\omega\rightarrow0}\lim_{\vecp\rightarrow 0}\sum_{i=1}^{3}\frac{\rho^{V}_{ii}(\omega,\vecp,T)}{\omega} ,
\label{eq:HQ_diffusion_formula}
\ee
where $\chi^{00}$ is the quark number susceptibility that is defined through the zeroth component of the temporal correlator 
in the vector channel.  The heavy quark diffusion coefficient is also related to the ratio of shear viscosity 
over entropy density $\eta/s$~\cite{Rapp:2008qc}.

On the lattice, however, spectral functions cannot be computed directly. The quantity that can be directly computed
is the Euclidean temporal correlation function $G_H(\tau,\vecp)$. It is defined as
\be
G_H(\tau,\vecp) =\int~\md^3\vecx~e^{-i\vecp\cdot\vecx}~\left <J_H(\tau,\vecx)J_{H}(0,\vecnull)\right > ,
\ee
where $G_{H}(\tau,\vecp)$ is the analytic continuation of $D^{+}(t,\vecp)$ from real to imaginary time
$G_H(\tau,\vecp)=D^{+}(-i\tau,\vecp)$.
By using the KMS relation and the above equation, one can relate the correlation function to the spectral function,
\be
G_H(\tau,\vecp) = \int_0^{\infty}\frac{\md\omega}{2\pi}\,\, \rho_H(\omega,\vecp,T)\, \,\,\frac{\cosh(\omega(\tau-1/2T))}{\sinh(\omega/2T)}.
\label{cor_spf_relation}
\ee
The above equation is the basic relation for extracting spectral functions from meson correlation functions calculated in the imaginary time formalism.
To extract the spectral function reliably one needs large number of points in the temporal direction of the lattice. Normally the Maximum Entropy Method (MEM) is used to 
extract spectral functions from correlators~\cite{Asakawa:2000tr}. As the temporal extent, $aN_{\tau}$, is restricted by the inverse temperature, in order to 
have larger number of points $N_\tau$ in the temporal direction one has to decrease lattice spacing $a$ at fixed values of temperature. As a consequence the computing cost will
increase. The Euclidean spatial correlation function $G_H(z,\vecp_\perp,p_\tau)$ can be computed via an integration over the Euclidean time $\tau$
and two spatial directions $\vecx_{\perp}=(x,y)$
\be
G_H(z,\vecp_{\perp},p_{\tau}) = \int^{1/T}_{0} \mathrm{d}\tau \int \mathrm{d} \vecx_{\perp}\, \exp(-i\tilde{\vecp}\cdot\tilde{\vecx}) \left\langle J_H(\tau,\vecx) J_H(0,\vecnull) \right\rangle,
\ee
where $\vecp_{\perp}=(p_{x},p_{y})$, $\tilde{\vecx}=(\vecx_{\perp},\tau)$ and $\tilde{\vecp} =( \vecp_{\perp}, p_\tau)$. The spatial correlation function
is related to the spectral function through
\be
G_H(z,\vecp_{\perp},p_{\tau}) = \int_{-\infty}^{\infty} \frac{\mathrm{d}p_z}{2\pi}\, \exp (ip_z\,z) \int_{0}^{\infty} \frac{\mathrm{d}\omega}{2\pi}\, \frac{2\omega}{\omega^2+p_\tau^2} \,\,\rho_H(\omega,\vecp_{\perp},p_z,p_{\tau}, T).
\ee
The above equation is much more complicated than that of the temporal correlation function and it thus is not so straightforward to extract the spectral function from the spatial correlation function.
However, one advantage of  spatial correlation functions over temporal correlation functions is that the physical extent in the spatial direction is not restricted by the temperature and thus by going to 
large distances one can extract the exponentially decayed constant, i.e. screening mass.

Bottomonium states are too heavy to be implemented in direct lattice QCD calculations. The lattice spacing required for this would need to be much smaller than the inverse of the bottom quark mass. 
A solution is provided by the Non-Relativistic QCD  (NRQCD), which is an effective theory of QCD where physics above the scale of the heavy quark mass $M$ is integrated out.
 In the framework of NRQCD, it is sufficient to have the lattice spacing $a$ as coarse as $M \gg 1/a \gg Mv$. Thus by construction there is no continuum limit of NRQCD on the lattice.
The correlation function in NRQCD is related to the spectral function through
\be
G_H(\tau,\vecp) = \int_0^{\infty}\frac{\md\omega}{2\pi}\, \rho_H(\omega,\vecp,T) \exp(-\omega \tau)~~~~~~~~~~~~~~~~~~~~~~(\rm NRQCD).
\label{cor_spf_relation_nrqcd}
\ee
Note that the integral kernel $\exp(-\omega\tau)$ has no temperature dependence and the correlation function $G_H(\tau,\vecp)$
in Eq.~(\ref{cor_spf_relation_nrqcd}) is not symmetric at $\tau=1/2T$. These are in contrast to $G_H(\tau,\vecp)$ defined in Eq.~(\ref{cor_spf_relation}).

\section{Chiral and axial symmetries}
\label{sec:basic_symmetry}

The QCD Lagrangian with $N_f$ massless quark flavors has a large group of symmetry
\be
SU(N_f)_L\, \otimes \,SU(N_f)_R \,\otimes\, U(1)_V\, \otimes\, U(1)_A .
\ee
In the vacuum the chiral symmetry $SU(N_f)_L\, \otimes \,SU(N_f)_R $ of QCD is spontaneously broken into
a smaller $SU(N_f)_V$ vacuum symmetry. It is expected that at sufficiently high temperature the QCD matter will transit from 
a chiral asymmetric phase to a chiral restored phase. The order parameter for such a phase transition is the chiral condensate
$\langle \bar{\psi}\psi\rangle$, which is nonzero below the chiral phase transition temperature $T_c$ and zero above $T_c$. The chiral 
symmetry restoration can also be signaled by the degeneracy of the vector and axial vector mesons, e.g. $\rho$ and $a_1$.

\begin{figure}[htp!]
\hspace{4cm}\resizebox{0.38\columnwidth}{!}{ \includegraphics{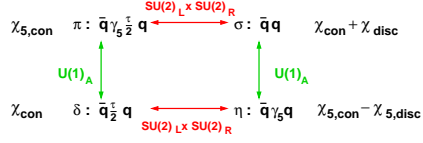}} 
\caption{Symmetry transformations relating pseudo scalar and scalar mesons in flavor non-singlet and singlet channels in $N_f=2$ QCD~\cite{HotQCD:2012ja}.}
\label{fig:symmetry}       
\end{figure}

The breaking of chiral symmetry $SU(3)_L\, \otimes \,SU(3)_R$  predicts the existence of nine massless Goldstone bosons. However, there are only eight light mesons in nature, i.e.
 octet mesons $\pi^{\pm}$, $\pi^0$, $K^{\pm}$, $K^0$, $\overline{K^0}$, $\eta$. The ninth ``Goldstone" boson, i.e. the singlet meson $\eta^{\prime}(958)$, has a much larger mass. Such a mass splitting
 is due to the violation of the axial $U(1)_A$ symmetry by the axial anomaly present at the quantum level~\cite{Adler:1969gk,Bell:1969ts}. The effective restoration of the $U(1)_A$ symmetry may lead to a softening of the $\eta^{\prime}$
 mass which would provide interesting experimental signatures.
 
 The degree of $U(1)_A$ symmetry breaking may affect our understanding of QCD phase diagram. In the chiral limit of two flavor QCD, the order of the chiral phase transition is expected to be second order
 belonging to the three dimensional O(4) universality class if the $U(1)_A$ is strongly broken near the phase transition. If the $U(1)_A$ symmetry breaking is negligible or $U(1)_A$ is restored near the phase transition
 then the chiral phase transition can become first order~\cite{Pisarski:1983ms}. Thus the nature of the chiral phase transition may depend critically on the strength of $U(1)_A$ symmetry breaking.
 
The way to investigate the chiral and $U(1)_A$ symmetry on the lattice is to look at various correlation functions and susceptibilities. The 
susceptibilities $\chi_{\pi,\delta,\sigma,\eta}$ are the four volume integration of the corresponding correlation functions. $\chi_{\rm disc}$ is the disconnected
part of the susceptibility of the chiral condensate $\langle\bar{\psi}\psi\rangle$ and it equals to the disconnected part of $\chi_{\sigma}$, and $\chi_{\rm 5,disc}$ is the disconnected part of the
susceptibility of $\langle\bar{\psi}\gamma_5\psi\rangle$ and it equals to disconnected part of $\chi_{\eta}$. They can also be written in terms of other susceptibilities 
as shown in Fig.~\ref{fig:symmetry} and thus the $U(1)_A$ and chiral  symmetries are restored through the following degeneracies:
\bea
U(1)_A \,\, {\rm symmetry}&:&  \quad \quad \chi_\pi = \chi_\delta \quad {\rm and}\quad \chi_{\delta} + \chi_{\rm disc} = \chi_{\pi} - \chi_{\rm 5,disc}\, , \\
SU(2)_L\times SU(2)_R \,\, {\rm symmetry}&:&   \quad \quad \chi_\pi = \chi_\delta + \chi_{\rm disc} \quad {\rm and}\quad \chi_{\delta} = \chi_{\pi} - \chi_{\rm 5,disc}\, .
\eea
We see that if $U(1)_A$ symmetry is restored the values of $\chi_\pi -\chi_\delta$ and $\chi_{\rm disc} + \chi_{\rm 5,disc}$ should both go to zero  and if chiral symmetry is restored
$\chi_\pi - \chi_\delta$ should equal to $\chi_{disc}$ and $\chi_{disc} -\chi_{5,disc}$ should be zero. When both chiral and $U(1)_A$ symmetries are restored,
$\chi_{\rm disc}$ and $\chi_{5,disc}$ should both equal to zero.

\section{Lattice QCD results}
\label{sec:results}

\subsection{Thermal dilepton rates, electrical conductivity and soft photon emission rate}
\label{sec:light}

\begin{figure}[htp!]
\hspace{2.5cm}\resizebox{0.3\columnwidth}{!}{ \includegraphics{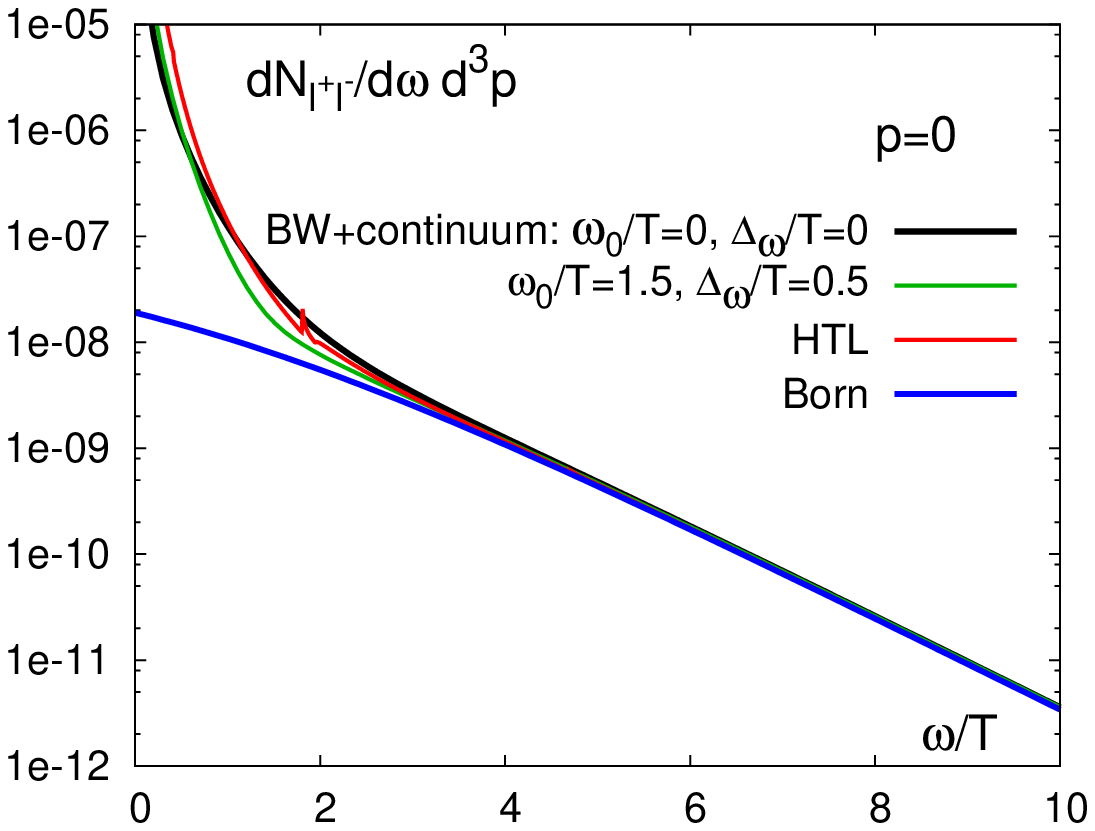}} ~~~~\resizebox{0.3\columnwidth}{!}{ \includegraphics{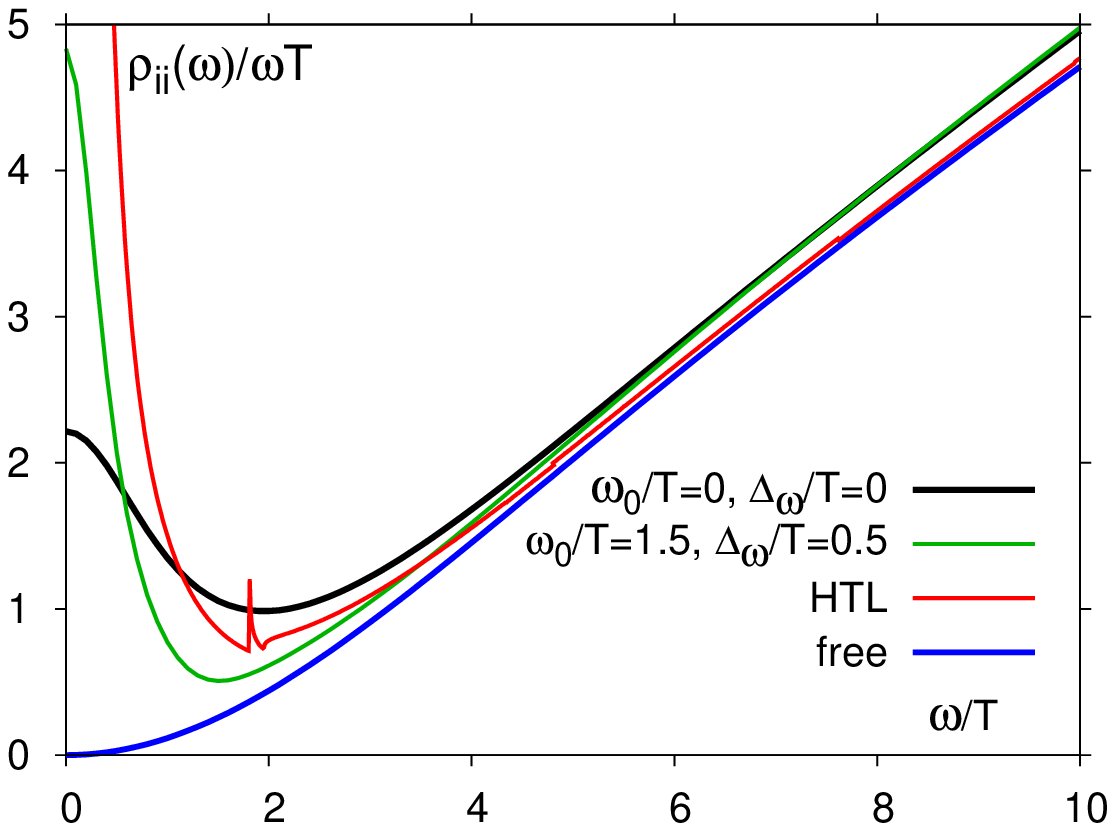}} 
\caption{Left: thermal dilepton rate in 2-flavor QCD. Shown are results from
fits using Eq.~(\ref{eq:ansatz}) without a cut-off on the continuum contribution ($\omega_0/T=0$) and
with the largest cut-off tolerable in the fit ansatz ($\omega_0/T=1.5$).
The HTL curve is for a thermal quark mass $m_T/T=1$ and the 
Born rate is obtained by using the free spectral function. Right: the spectral functions that entered the calculation of the dilepton rate~\cite{Ding:2010ga}.}
\label{fig:dilepton}       
\end{figure}

Recently a detailed calculation of the light vector correlation function on large isotropic quenched lattice was reported at a fixed value of temperature 
$T\simeq 1.45~T_c$~\cite{Ding:2010ga}. Calculations have been performed at several lattice spacings and volumes.
This enables a reliable extrapolation of the vector correlation function to the continuum limit 
for a large Euclidean time interval $0.2 \le \tau T\le 0.5$. From the comparison of the free correlation function and the continuum extrapolated vector correlation function at $T\simeq 1.45~T_c$, it was found
that the high energy part of the spectral function at $T\simeq1.45T_c$ is quite close to the free field theory, thus the following ansatz for the spectral function 
is used to analyze the continuum extrapolated vector correlation function

\be
\rho_{ii} (\omega) =
2\chi_q c_{BW}   \frac{\omega \Gamma/2}{ \omega^2+(\Gamma/2)^2} + {3 \over 2 \pi} \left( 1 + k \right) 
\,\omega^2  \,\tanh (\omega/4T)\,\Theta(\omega_0,\Delta_\omega) 
\, ,
\label{eq:ansatz}
\ee
where $c_{BW}$ and $\Gamma$ are fit parameters to adjust the Breit-Wigner (BW) term, and $k$ is a fit parameter accounting for corrections to the free
field theory in the second term. To understand the systematic errors that arise from a specific fit ansatz, a function $\Theta(\omega_0,\Delta_\omega) =
\left( 1+{\rm e}^{(\omega_0^2-\omega^2)/\omega\Delta_\omega} \right)^{-1}$ is also adopted in the second term of Eq.~(\ref{eq:ansatz}).
In the limit of $\Delta_\omega\rightarrow 0$ the function $\Theta(\omega_0,\Delta_\omega)$ becomes a step function with discontinuity 
at $\omega_0$. Fits with two extreme cases of parameter sets of $\omega_0$ and $\Delta_\omega$ that are constrained by data are shown in
Fig.~\ref{fig:dilepton}. In the left panel of Fig.~\ref{fig:dilepton} the thermal dilepton rates are compared to 
those calculated within the hard thermal loop (HTL) approximation~\cite{Braaten90} using a thermal quark mass $m_T/T=1$. Obviously they 
are in good agreement for $\omega/T\gtrsim 2$. For $1\lesssim \omega/T\lesssim 2$ 
differences between the HTL spectral function and  
lattice results is about a factor of two, which also is the intrinsic
uncertainty in the spectral analysis. At energies $\omega/T\lesssim1$ 
the HTL results grow too rapidly, as is well known. As the results shown in Fig.~\ref{fig:dilepton} is only for one single temperature,
it is important to have results from lattice QCD at various temperatures around $T_c$ in the near future.

The vector spectral function is shown in the right panel of Fig.~\ref{fig:dilepton}. There is no resonance peak
seen in the spectral function. Through Eq.~(\ref{conduct}) the slope of the spectral function at vanishing energy gives an estimate for the electrical conductivity and the zero energy limit of the thermal 
photon rate~\cite{Ding:2010ga}
\begin{equation}
1/3 \ \lesssim \ \frac{1}{C_{em}}
\frac{\sigma}{T} \ \lesssim \ 1, \quad  \lim_{\omega \rightarrow 0} \omega \frac{{\rm d} R_\gamma}{{\rm d}^3p} =
\left( 0.0004 \ -\ 0.0013 \right) T_c^2 \quad  {\rm at}  \quad T\simeq 1.45\ T_c \; .
\label{range}
\ee
Different values of the electrical conductivity  from lattice QCD calculations were reported in Ref.~\cite{Gupta:2003zh,Aarts:2007wj}, however, these values,
unlike those reported in Ref.~\cite{Ding:2010ga}, are unrenormalized and not extrapolated to the continuum limit.
The temperature dependence of the electrical conductivity has also been studied on the lattice recently. It was found that $\sigma/T\approx 0.33 C_{em}$
 and is almost independent on temperature in the temperature range $1.16 T_c \lesssim T \lesssim2.98 T_c$~\cite{Francis:2011bt} . 
However, the results reported in Ref.~\cite{Francis:2011bt} were obtained at only one value of the lattice
cutoff and were not extrapolated to the continuum limit. Nevertheless these results are consistent with those in Ref.~\cite{Ding:2010ga}.
The leading order pQCD calculations give a much larger result, i.e. $\sigma/T\approx 6~e^2$~\cite{Arnold:2003zc}.
Note also that the electrical conductivity obtained  from lattice QCD is more than an order of magnitude larger than the that  
 of a pion gas above $T_c$~\cite{FernandezFraile:2009mi}.

\subsection{Heavy quarkonia correlators and spectral functions}

The most direct approach to check whether a quarkonium state survives in the hot medium is to investigate the temperature dependence
of its spectral function. Recently a detailed analysis of the charmonium temporal correlation and spectral functions computed on fine and large isotropic
quenched lattices at $T=0.73,~1.46,~2.20$ and $2.93~T_c$ was reported in Ref.~\cite{Ding:2012sp}.  Computing time can be saved by doing simulations on anisotropic lattice but the 
lattice cutoff effects are much more severe in this case as discussed in Ref.~\cite{Ding:2012sp}. This work is mainly based on the MEM to extract the spectral functions. 
As the number of data points in the temporal correlation function is very important in this type of 
study, an enormous computing effort has been made to have large isotropic lattices, e.g. $128^3\times$ 48 at 1.46 $T_c$.
The uncertainties on the spectral function from MEM have been carefully studied. The results for the charmonium spectral function in
the vector channel is shown in the left panel of Fig.~\ref{fig:charm}. It is obvious that at $1.46~T_c$ the ground state peak seen at $0.73~T_c$ disappears, i.e.
$J/\psi$ does not survive at $T\geqslant1.46~T_c$. $\eta_c$, $\chi_{c0}$ and $\chi_{c1}$ are also found to disappear at $T\geqslant1.46~T_c$~\cite{Ding:2012sp}.
The fate of S wave sates ($J/\psi$ and $\eta_c$) suggested by the analysis in Ref.~\cite{Ding:2012sp}
is different from previous lattice QCD studies that suggested the survival of S wave sates up to 2$~T_c$. This is mainly due to the fact that most of the previous lattices studies 
were done on anisotropic lattices and had a fewer number of points in the temporal direction~\cite{Ding:2012sp}. On the other hand, the results for P wave states ($\chi_{c0}$ 
and $\chi_{c1}$) from this work are compatible with previous results,  i.e. they does not survive at $T\geqslant1.46~T_c$. However, these results from Ref.~\cite{Ding:2012sp} do not necessarily mean that all the charmonium states
melt at the same temperature. One needs further lattice QCD computations at $T_c<T<1.46T_c$ to locate the dissociation temperatures of charmonium states.

\begin{figure}[htp!]
\hspace{2.5cm}\resizebox{0.28\columnwidth}{!}{ \includegraphics{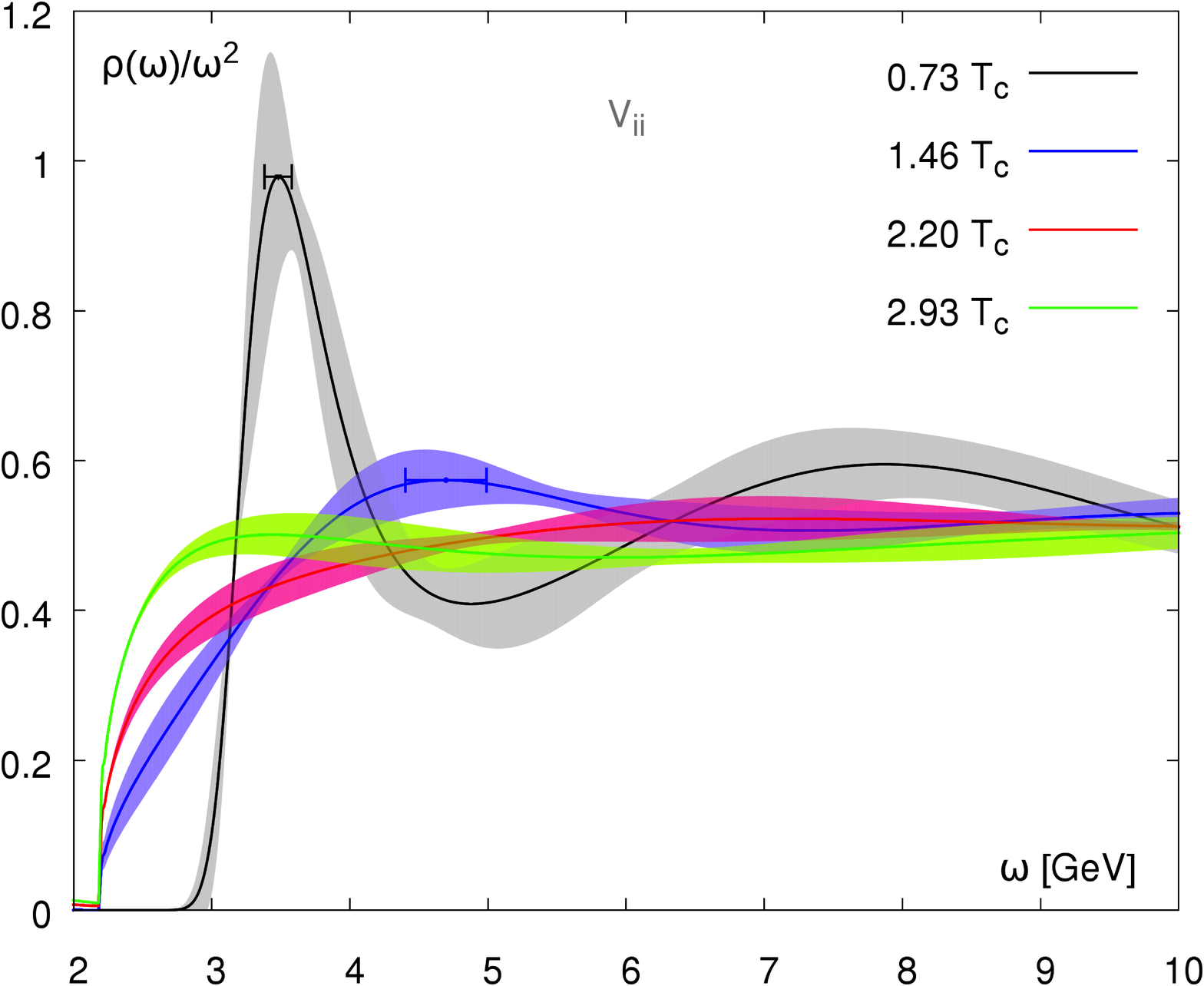}} ~~~\resizebox{0.3\columnwidth}{!}{ \includegraphics{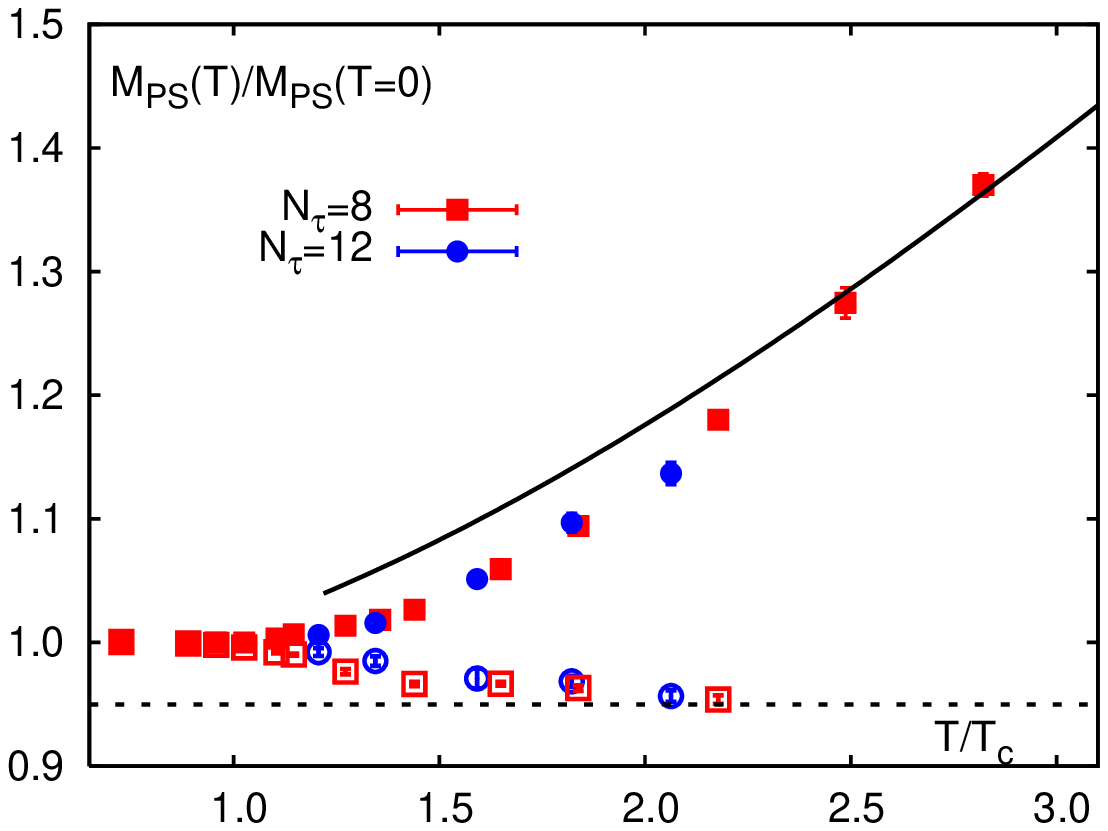}} 
\caption{Left: Statistical uncertainties of output spectral functions in the $V_{ii}$
channel at  temperatures $T=0.73,~1.46,~2.20,~2.93T_c$. The shaded areas are statistical errors of amplitudes of output spectral
functions from Jackknife analyses and the solid lines inside the shaded areas are mean values
of spectral functions. The horizontal error bars at the first peaks of spectral functions at $0.73~T_c$ and $1.46~T_c$ stand for the statistical uncertainties of the peak location obtained from 
Jackknife analyses~\cite{Ding:2012sp}. Right: Pseudo scalar charmonium screening
masses calculated with antiperiodic (filled symbols) and periodic
(open symbols) boundary conditions as function of the temperature. 
Solid (dashed) lines correspond to the free theory
prediction with charm quark mass $m_c=1.42$ GeV for antiperiodic (periodic)
boundary conditions~\cite{Karsch:2012na}.}
\label{fig:charm}       
\end{figure}

Spatial charmonium correlators and screening masses have also been calculated on the lattice in full QCD~\cite{Karsch:2012na}. As shown in the right panel of Fig.~\ref{fig:charm},
at $T\gtrsim1.5~T_c$ the screening mass in the pseudo scalar channel differs strongly from its zero temperature value and it depends strongly on the boundary conditions. All this suggests
the dissociation of bound states in the pseudo scalar channel at $T\gtrsim1.5~T_c$. This finding is consistent with that from the study of charmonium spectral functions extracted from temporal correlation
functions reported in Ref.~\cite{Ding:2012sp}.
\begin{figure}[htp!]
\hspace{3.5cm}\resizebox{0.4\columnwidth}{!}{ \includegraphics{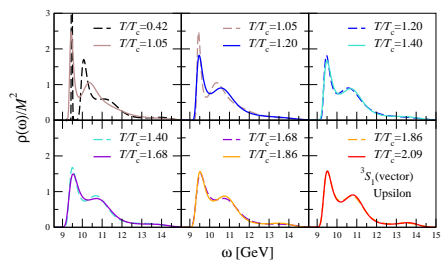}} 
\caption{Bottomonium spectral functions in the vector channel from Non-Relativisitic QCD calculations~\cite{Aarts:2011sm}.}
\label{fig:bottom}       
\end{figure}

For the fate of bottomonia at finite temperature, a lattice calculation in the framework of NRQCD has  been performed  recently~\cite{Aarts:2011sm}.
Lattice QCD simulations were performed with two light quarks on highly anisotropic lattice using NRQCD for bottom quarks.
The spectral functions shown in Fig.~\ref{fig:bottom} are extracted from correlations functions using the MEM. It can clearly be seen from that figure that 
the ground state in the $\Upsilon$ channel survives up to $2.09~T_c$. Results for the $\eta_{b}$ channel are similar.
The calculation has also been extrapolated exploringly to nonzero baryon density~\cite{Hands:2012yy}. However, one has to note 
that there is no continuum limit for the lattice version of NRQCD and the calculations in Ref.~\cite{Aarts:2011sm,Hands:2012yy} were performed on highly anisotropic lattices that can distort the physics.

The properties of heavy quarkonia in the vacuum have been studied based on a non relativistic Schr\"odinger equation using either modeled or lattice QCD computed 
heavy quark potentials. However, the definition of the heavy quark potential (free energy or internal energy ) is ambiguous at nonzero temperature.
Recently it has been proposed by authors in Ref.~\cite{Rothkopf:2011db} that in the nonrelativistic limit the leading order potential in an effective Schr\"odinger equation can be 
appropriately defined by assuming that the spectral density of the Wilson loop has peaks according to Breit-Wigner distributions. 
In this framework, the peak location of the spectral density corresponds to the real part of the heavy quark potential while the peak width corresponds
to the imaginary part of the heavy quark potential. Using the MEM it is feasible to extract the peak location, however, it is very hard to 
extract the width of the peak reliably.

The charm diffusion coefficient obtained from the slope of the vector spectral function is almost $1/\pi T$ at $T\in(1.46~T_c, 2.93~T_c)$ and compatible with zero 
below $T_c$~\cite{Ding:2012sp}. The mean value of the heavy quark diffusion coefficient obtained from the computation of heavy quark effective theory on the lattice is somewhat 
larger~\cite{Francis:2011gc,Banerjee:2011ra}, i.e. $~2/\pi T$ but compatible
with the charm diffusion coefficient from Ref.~\cite{Ding:2012sp} within errors. These values from lattice QCD
are smaller than that from pQCD calculations in the investigated temperature window~\cite{CaronHuot:2007gq} and lies in the range used in phenomenological
models~\cite{Gossiaux:2012th}.

\subsection{Chiral phase transition and $U(1)_{A}$ symmetry restoration}

Studies on this topic have also been carried out recently in Ref.~\cite{Cheng:2010fe}. In this study the improved (p4) staggered action was used and the effective restoration of the 
$U(1)_A$ symmetry occurs at a temperature larger than the QCD transition temperature indicated from the degeneracy of the screening masses of pseudo scalar and scalar correlators.
A better discretized version of staggered fermions has also been implemented in Ref.~\cite{Ohno:2011yr} to investigate the fate of the axial symmetry at finite temperature and their results
suggests that $U(1)_A$ symmetry still remains broken at about 1.1 $T_c$. Due to subtle issues concerning the realization of chiral and axial symmetries in the staggered discretized fermion action, 
better discretization 
schemes, e.g. domain wall fermions and overlap fermions, which preserve the full chiral symmetry of continuum QCD and reproduce the correct axial anomaly even at nonzero lattice spacing,
have also been implemented to study this topic.
\begin{figure}[htp!]
\hspace{2.5cm}\resizebox{0.25\columnwidth}{!}{ \includegraphics{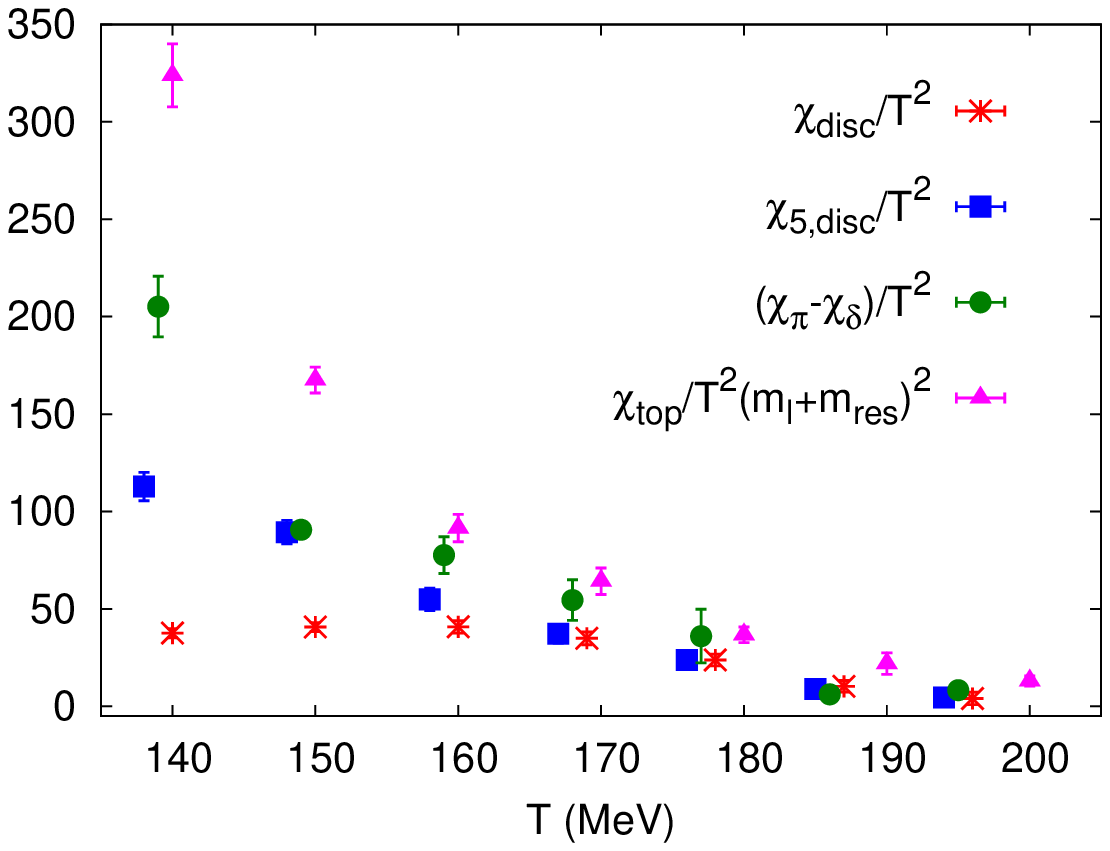}}~\resizebox{0.28\columnwidth}{!}{ \includegraphics{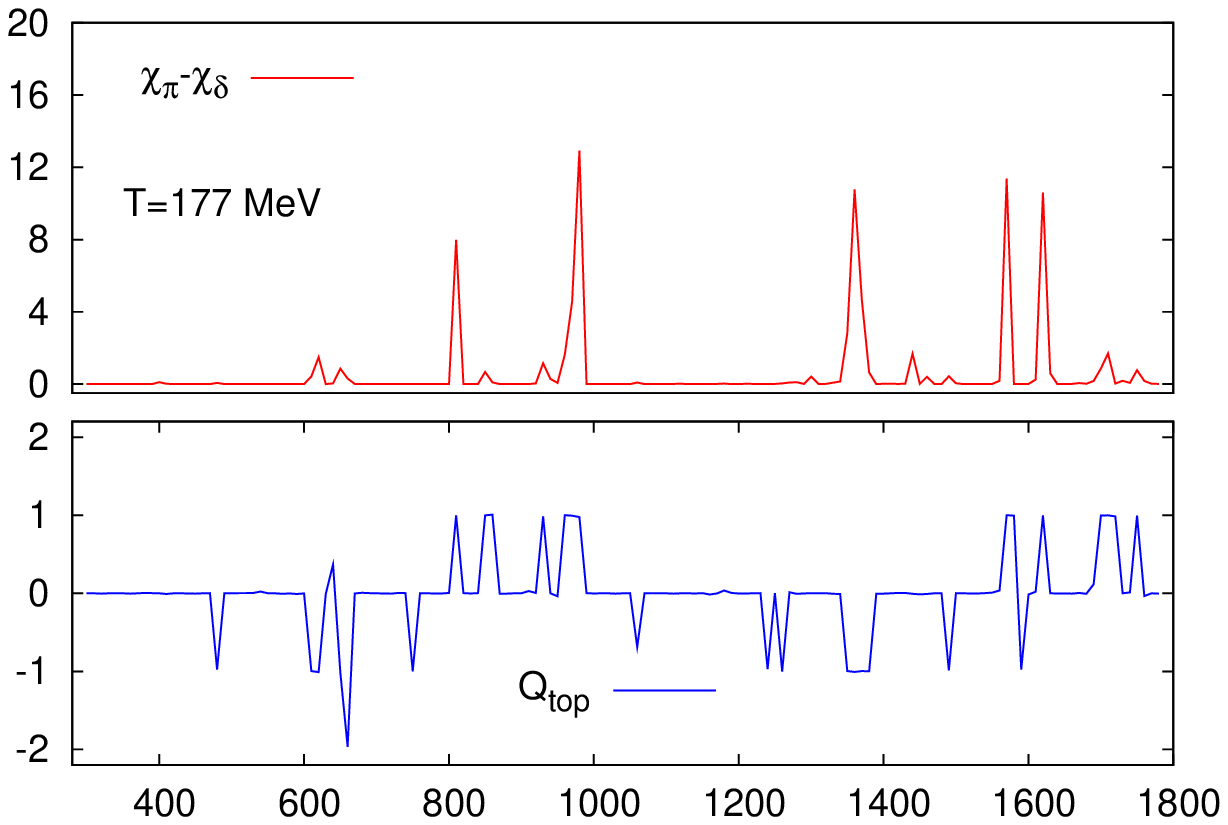}}
\caption{Left:  susceptibilities as function of temperature, Right: the time histories of the topological charge $Q_{top}$ and  $\chi_\pi - \chi_\delta$.
 Figures are taken from Ref.~\cite{HotQCD:2012ja}.}
\label{fig:sus}       
\end{figure}

Using domain wall fermions the HotQCD collaboration  recently investigated the restoration of the spontaneously broken chiral symmetry and the effective
restoration of the anomalously broken $U(1)_A$ symmetry at finite temperature  on $16^3\times8$ lattices with $m_\pi\approx200$ MeV~\cite{HotQCD:2012ja}.
One of the main results is shown in Fig.~\ref{fig:sus}. The three quantities, $\chi_{\rm disc}$, $\chi_{\rm 5,disc}$ and $\chi_\pi-\chi_\delta$,
agree within errors for temperature $T\gtrsim 170$ MeV. This suggests the restoration of the $SU(2)_L\times SU(2)_R$ symmetry. The $U(1)_A$ symmetry, 
on the other hand, remains broken up to temperature $T\approx 200$ MeV as indicated from the nonzero value of $\chi_{\pi}-\chi_{\delta}$. If 170 MeV is taken as the pseudo 
critical transition temperature, this indicates that $U(1)_A$ still remains broken at $T\approx 1.2 T_{pc}$. 
As shown in the right panel of Fig.~\ref{fig:sus}, the nonzero value of $\chi_\pi - \chi_\delta$ is related to the nonzero topological charge $Q_{top}$, which will be suppressed 
in the large volume limit. To draw reliable conclusion on $U(1)_A$ symmetry restoration at high temperature, one needs to carry out simulations with larger volumes and perform
detailed analysis on the Dirac operator eigenvalue spectrum.
The JLQCD collaboration using overlap fermions performed simulations at
higher temperatures mainly on $16^3\times8$ lattices with $m_\pi\gtrsim290~$MeV and $Q_{top}$ fixed to be zero~\cite{Cossu:2012gm}. 
They found that $U(1)_A$ symmetry is restored at $209$ MeV by looking at the degeneracies
of correlators and the gap around origin in the Dirac operator eigenvalue spectrum. However, the volume effects on topological charge fixing might be large and 
need to be checked.

From the studies discussed above,  the effective $U(1)_A$ symmetry restoration seems to occur after the chiral symmetry restoration and $U(1)_A$ restoration temperature certainly needs further
studies. Of course, it would be very interesting to have good signals to probe the possible restoration of $U(1)_A$ symmetry in the heavy ion experiments~{\cite{Huang:1995fc,Kapusta:1995ww,Csorgo:2009pa}.

\section{Summary}
\label{sec:sum}

The lattice QCD calculations presented here show that there has been significant progress in our understanding
of strongly interacting matter through the in-medium hadron properties in recent years. However, more needs to be done
on the lattice to help to understand the heavy ion experiment data. The $U(1)_A$ symmetry restoration that is being intensively studied on the lattice may
be  observed in the heavy ion experiments from some interesting signals, e.g. a softening of the $\eta^{\prime}$ mass.

\section*{Acknowledgements}
The work is supported under contract
No. DE-AC02-98CH10886 with the U.S. Department of Energy.
HTD thanks Prasad Hegde and Swagato Mukherjee for useful discussions and Frithjof Karsch for critical reading of the manuscript.


\end{document}